\documentclass[twocolumn,prl,showpacs]{revtex4}

\usepackage{epsfig}
\usepackage{graphicx}
\usepackage{dcolumn}
\usepackage{bm}

\usepackage{epstopdf}
\newcommand{\mean}[1]{\langle #1 \rangle}

\begin{document}


\title{Rotational dynamics and friction in double-walled carbon nanotubes}

\author{J. Servantie}
\author{P. Gaspard}
\affiliation{Center for Nonlinear Phenomena and Complex Systems\\
Universit\'e Libre de Bruxelles, Code Postal 231, Campus Plaine, 1050
Brussels, Belgium.}

\date{\today}

\begin{abstract}
We report a study of the rotational dynamics in double-walled
nanotubes using molecular dynamics simulations and a simple
analytical model reproducing very well the observations. We show that
the dynamic friction is linear in the angular velocity for a wide
range of values. The molecular dynamics simulations show that for
large enough systems the relaxation time takes a constant value
depending only on the interlayer spacing and temperature. Moreover, the friction force increases linearly with contact area, and the relaxation time decreases with the temperature with a power law of exponent $-1.53 \pm 0.04$.
\end{abstract}

\pacs{68.35.Af; 85.35.Kt}

\maketitle

Since the pioneering work of Iijima on nanotubes \cite{Iijima}, research on their electronic and mechanic properties has been rapidly growing. Very high electric and thermal conductivities were observed and the nanotubes were shown to be as stiff as diamond in their axial direction with Young modulus of the order of TPa. These observations make the nanotubes promising candidates for nano-electrical and nano-mechanical devices. The experiments on the translational motion in multi-walled nanotubes of Cumings and Zettl \cite{Zettl} and Yu {\it et al.} \cite{Ruoff} motivated the study of these possible oscillators. 
Many numerical studies were performed to characterize the translational motion of double-walled nanotubes \cite{Servantie2003,Servantie2006,Jiang3,Zheng,Tangney}. On the other hand, experiments were also carried out to study the rotational motion of concentric nanotubes. Indeed, Fennimore {\it et al.} \cite{Fennimore} and Bourlon {\it et al.} \cite{Bourlon} built actuators based on multi-walled nanotubes. As for the translational oscillatory motion, no wear or fatigue was observed in the rotation. Recently, Kr{\'al} and Sadeghpour \cite{Kral} proposed to spin nanotubes with circularly polarized light, further motivating 
the use of nanotubes as possible axis of rotation at the nanoscale. However, few systematic analytical and numerical studies of rotation have been done, except the work of Zhang {\it et al.} \cite{Ruoff2004} where nonequilibrium molecular dynamics simulations were carried out to calculate the energy dissipation rate during the rotational motion in the presence of external thermostat.  Nevertheless, the friction between two carbon nanotubes which are rotating one with respect to the other has not yet been understood and characterized.

The purpose of the present Letter is to study the rotational dynamics of a double-walled nanotube (DWNT) by molecular dynamics simulations with the conservative Hamiltonian mechanics in order to avoid additional sources of dissipation and precisely characterize the rotational friction.  Since friction is related to the fluctuations of the many vibrational degrees of freedom, we set up a Langevin-type stochastic model and validate it by molecular dynamics simulations of typical DWNTs. This shows that dynamic
friction is proportional to the angular velocity up to high values, increases linearly with contact area, and as a power law of exponent $1.53 \pm 0.04$ with temperature.

We consider a DWNT with free ends in vacuum.  The Hamiltonian of the system of two concentric nanotubes is the same as in our previous
studies for the translational motion \cite{Servantie2003,Servantie2006}. The intratube interactions are modeled by the Tersoff-Brenner potential with the set of parameters of Table III in Ref. \cite{Brenner} and the intertube interactions by a 6-12 Lennard-Jones potential used for carbon material simulations \cite{Lu}. The DWNT is long enough so that the inner and outer nanotubes essentially rotate around a common axis with the angular velocities $\omega_1$ and $\omega_2$, respectively. Their angular momenta are given by $L_a=I_a\omega_a$ in terms of the moments of inertia $I_a$ ($a=1,2$). The equations of motion of this model are given by
\begin{equation}
\frac{dL_a}{dt} = I_a \frac{d\omega_a}{dt} = N_a \quad (a=1,2)
\label{eqs}
\end{equation}
with the torques $N_1=-N_2$ in order to conserve the total angular momentum $L_0=L_1+L_2$. We can introduce the angular velocity of the
center of inertia $\Omega=(I_1\omega_1+I_2\omega_2)/(I_1+I_2)=L_0/(I_1+I_2)$ and the relative angular velocity $\omega=\omega_1-\omega_2$ so that the equations of motion become $d\omega/dt = N_1/I$ and $d\Omega/dt=0$ with the relative moment of inertia $I=I_1 I_2 /(I_1+I_2)$. The total rotational kinetic energy of the nanotubes is given by
\begin{equation}
T_{\rm r}=\frac{1}{2}I_1 \omega_1^2+\frac{1}{2}I_2 \omega_2^2
= \frac{1}{2}(I_1+I_2) \Omega^2+\frac{1}{2}I \omega^2
\label{T_r}
\end{equation}
The corrugation against the rotational motion is very small compared to the thermal energy. Indeed, for the two DWNTs considered
in this Letter, namely (4,4)@(9,9) and (7,0)@(9,9), and the 6-12 Lennard-Jones potential we use \cite{Lu}, we find a corrugation
against rotation of respectively 0.00992 meV/atom and 0.00548 meV/atom. These values are of the same order as the ones of Zhang
{\it et al.} who used the Kolmogorov and Crespi potential \cite{Crespi}. Even for temperatures as low as 100 K (8.6 meV), thermal energy is much larger than corrugation energy against rotation. This observation allows us to neglect the mean static torque due to the corrugation in the equation of motion of the system. However, in order to model the rotational dynamics of the DWNT one should take into account dynamic friction.  If the angular velocity is small enough, we expect that the friction torque is linear in the angular velocity in analogy with the case of the translational motion \cite{Servantie2003,Servantie2006}. Indeed, the rotation of a nanotube inside another one is comparable to the the sliding friction of Xenon film on a silver surface considered by Daly and Krim \cite{Krim1996} and Tomassone {\it et al.} \cite{Krim1997} who showed that, in their system, friction is dominated by phonon
excitations and that dynamic friction is proportional to the sliding velocity. Accordingly, we model our system by a Langevin-type
equation without potential,
\begin{equation}
I\; \frac{d\omega}{dt}=-\chi\, \omega+N_{\rm fluct}(t)
\label{StochRot}
\end{equation}
where $\chi$ is a damping coefficient and $N_{\rm fluct}(t)$ a fluctuating torque of zero mean value and correlation function
$\langle N_{\rm fluct}(t) N_{\rm fluct}(t')\rangle \simeq 2 \chi \, k_{\rm B}T\, \delta(t-t')$ in terms of the temperature $T$ and Boltzmann's constant $k_{\rm B}$. This is justified because the correlation time is of the order of the inverse of the Debye frequency (about $50$ fs) as for the translational motion \cite{Servantie2003}. Since we observe relaxation times of the order of the nanosecond, one can assume this Markovian form for the evolution equation. We notice that the variations of the moments of inertia are neglected in Eq.(\ref{StochRot}).

A molecular dynamics simulation can be carried out by giving an initial angular momentum $L_0=I_1 \omega_0$ to the inner tube ($a=1$) and a zero one to the outer tube. After the relaxation time $\tau=I/\chi$, the system reaches a state of equilibrium in which both tubes rotate with the same angular frequency, $\langle\omega_1\rangle_{\rm eq}=\langle \omega_2\rangle_{\rm eq}$.  This
numerical observation is in agreement with the consequence of Eq. (\ref{StochRot}) that the mean relative angular velocity vanishes at
equilibrium. Using the conservation of total angular momentum, the equilibrium values of the angular momenta are thus given by $\langle
L_{a}\rangle_{\rm eq}=L_0 I_a/(I_1+I_2)$ for $a=1,2$. At equilibrium, the fluctuations of the angular velocity have the variance
$\mean{\omega^2}_{\rm eq}=k_{\rm B} T/I$.  For times longer than the relaxation $t \gg I/\chi$, the relative angle $\theta$ performs a
diffusive random rotation
\begin{equation}
\mean{\theta^2(t)} \simeq 2D(t-\tau)
\label{MSD}
\end{equation}
with the diffusion coefficient $D=k_{\rm B} T/\chi$. Moreover, the angular velocity autocorrelation function is given by
\begin{equation}
\mean{\omega(t)\omega(t')}_{\rm eq}=\frac{k_{\rm B} T}{I}e^{-\lambda
(t-t')}
\label{ACF}
\end{equation}
where we defined $\lambda=\chi/I=1/\tau$. If a relative angular velocity $\omega_0$ is initially given to the system, Eq. (\ref{StochRot}) shows that it decays on average according to
\begin{equation}
\mean{\omega(t)}=\omega_0 \; e^{-\lambda t}
\label{evolO}
\end{equation}
and the total rotational kinetic energy (\ref{T_r}) as
\begin{equation}
\mean{T_{\rm r}(t)}=\frac{T_0}{I_1+I_2}\left(I_2e^{-2\lambda
t}+I_1\right)+\frac{k_{\rm B} T}{2}\left(1-e^{-2\lambda t}\right)
\label{evolT}
\end{equation}
where $T_0=\mean{T_{\rm r}(0)}$ denotes its initial
value. Hence, we conclude that,  in the long-time limit $t 
\rightarrow \infty$,
the rotational kinetic energy reaches the value
\begin{equation}
\mean{T_{\rm r}(\infty)}=\frac{L_0^2}{2(I_1+I_2)}+\frac{k_{\rm B} T}{2}
\label{TRINFTY}
\end{equation}


We now show the results of the molecular dynamics simulations. The
integration is done with a velocity Verlet scheme of time step 2 fs
for temperatures up to 600 K and 1 fs for larger temperatures.
Angular velocity is fixed by giving a tangential velocity
corresponding to $\omega_0$ to the velocities of each atom of the
inner tube. We compute the average relative angular velocity
$\langle\omega\rangle$ and rotational kinetic energy $\langle T_{\rm
r}\rangle$ from the angular momenta and moments of inertia in the
axial direction. The angle is then obtained by the time integral of
the angular velocity.

As a first test, we estimated the maximum
angular velocity the nanotube could endure before its dislocation.
One can easily understand that the velocity of rotation should be
smaller than the radial phonon velocity to guarantee the stability of
the system. Indeed,  we have calculated the critical angular velocity
at which the nanotube collapses for increasing radii of the nanotube.
We found out that the critical velocity $v_{\rm c}=\omega_{\rm c} R$
is constant and approximately equal to $7.98 \pm 0.80$ km/s, close to
the velocity 8 km/s of radial phonons \cite{Servantie2006}. However,
Zhang {\it et al.} \cite{Ruoff2004} observed the collapse of their
nanotube for angular velocities higher than 4.7 rad/ps, corresponding
to a radial velocity of approximately 3.1 km/s. This smaller value is
probably due to the fact that the angular velocity was applied
incrementaly to constraint atoms at the end of the tube while we give
as an initial condition the angular velocity to all the atoms of the
nanotube.

\begin{figure}[h]
\includegraphics[scale=0.6]{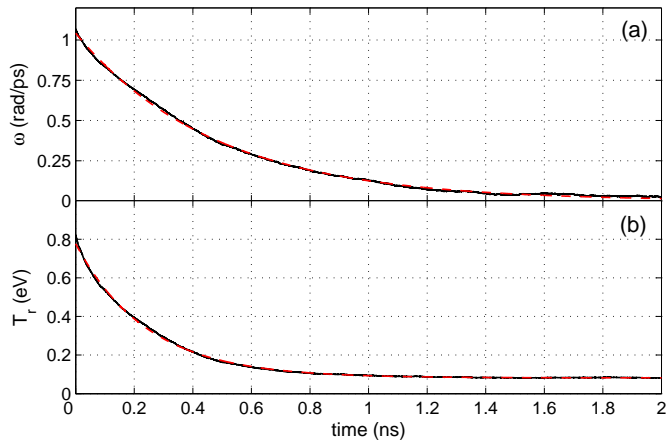}
\caption{(Color online) Plots of the time evolution of (a) the average relative
angular velocity $\langle\omega\rangle$ and (b) the average
rotational kinetic energy $\langle T_{\rm r}\rangle$ for the
armchair-armchair DWNT described in the text.
The dashed lines are
the fits by (a) Eq. (\ref{evolO}) and (b) Eq. (\ref{evolT}).
The averages are computed with about $10^3$ trajectories.}
\label{figEvol}
\end{figure}

\begin{figure}[h]
\includegraphics[scale=0.6]{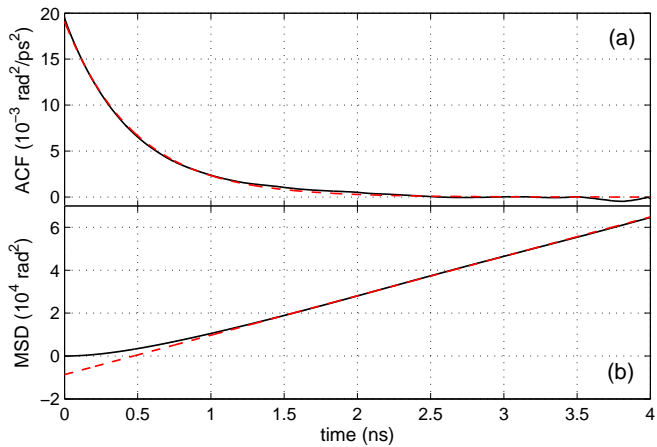}
\caption{(Color online) Plots of the time evolution of (a) the velocity
autocorrelation function and (b) the mean square displacement of the
angle for the armchair-armchair DWNT described in the text. The
dashed lines are the fits by (a) Eq. (\ref{ACF}) and (b) Eq.
(\ref{MSD}).
The statistics are computed with about $10^3$ trajectories.}
\label{figDiff}
\end{figure}

We apply our model to the armchair-armchair DWNT (4,4)@(9,9) with
both nanotubes of length 2.11 nm and inertia moments $I_1=139.11$ amu
${\rm nm}^2$ and $I_2=1547.9$ amu ${\rm nm}^2$ ($I=127.64$ amu ${\rm
nm}^2$) at room temperature (300 K). We depict in Fig. \ref{figEvol}
the time evolution of the angular frequency and the rotational
kinetic energy from an initial relative angular velocity of
$\omega_0=1.07$ rad/ps. The relaxation time calculated from the angular
velocity in Fig. \ref{figEvol}(a) is 473 ps while the relaxation time
from the energy is 243 ps which is in very good
agreement with the factor 2 appearing between 
the decay rates in 
Eqs. (\ref{evolO}) and (\ref{evolT}).  The decay
of the rotational kinetic energy is well fitted
by Eq. (\ref{evolT}) with the initial kinetic energy $T_0=0.822$ eV.
On the other hand, the simulations show that the equilibrium
kinetic energy takes the value 0.0816 eV, which is in excellent
agreement with Eq. (\ref{TRINFTY}) of the model.
In order to confirm
the validity of the stochastic model described in this Letter, we use
the same code,
but with a vanishing initial angular velocity. The
velocity autocorrelation function and the mean square displacement
are calculated after a time of 1 ns longer than the relaxation. The
results are depicted in Fig. \ref{figDiff}. The velocity
autocorrelation function is well fitted by the exponential
(\ref{ACF}) with a relaxation time of 475.9 ps while the linear fit
(\ref{MSD}) on the mean square displacement gives $\tau=$ 471.3 ps.
These results are in excellent agreement with the previous
simulations. The small differences are due to the statistical error,
and can be reduced with longer simulation times.
These simulations successfully demonstrate that the dynamic friction 
against the
rotational motion is proportional to angular velocity in DWNTs.

\begin{figure}[h]
\includegraphics[scale=0.6]{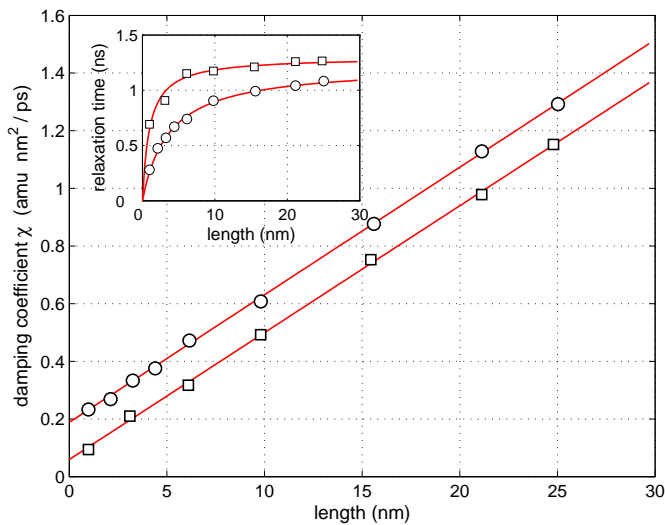}
\caption{(Color online) Dependence of the damping coefficient $\chi$ on length. The
circles and squares are the results of the simulations for
respectively the armchair-armchair DWNT (4,4)@(9,9) and the
zigzag-armchair DWNT (7,0)@(9,9). The linear fits have the same slope
0.044 amu nm/ps. The inset is the relaxation time versus length
$\ell$, which is well fitted by
$\tau=\tau_{\infty}\ell/(\ell+\ell_0)$.
$\tau_{\infty}= 1.22$ ns and
$\ell_0=3.6$ nm for the armchair-armchair DWNT.
$\tau_{\infty}=1.30$
ns and $\ell_0=1.0$ nm for the zigzag-armchair DWNT.}
\label{figLeng}
\end{figure}

We calculated the dependence of the damping coefficient on contact
area at room temperature (300 K) for increasing lengths of our DWNTs.
In order to obtain this result, we measured the relaxation time
$\tau$ of the angular velocity and calculated the damping coefficient
as $\chi=I/\tau$. The damping coefficient $\chi$ is depicted as a
function of the length of the tubes in Fig. \ref{figLeng} where we
see that the linear fit is in excellent agreement with the
simulations. The friction force does not vanish at zero contact area
because of the attractive forces between the surfaces \cite{JKR}.
Having obtained the dependence on length, we investigated the
dependence on radius. We calculated the relaxation time for
increasing radii of zigzag-armchair DWNTs. Our numerical simulation
gives no systematic dependence of the relaxation time on radius
except the variations due to the different interlayer spacings
considered. Hence the dependence on radius of the damping coefficient
comes only from the inertia moment $I$. The inertia moment of a
nanotube of length $\ell$, radius $R$ and surface mass density
$\sigma$ is very well approximated by $I\simeq 2\pi\sigma \ell R^3$.
For the hexagonal lattice, the surface mass density is $\sigma=4
m/(3\sqrt{3}a_{\rm CC}^2)$ where $m=12$ amu is the atomic mass of
carbon and $a_{\rm CC}=1.42$ \AA\ the carbon-carbon bond length.
Hence, $\sigma=4.55 \; {\rm amu}/\mbox{\AA}^2$. In order to recover
the experimental conditions \cite{Fennimore,Bourlon} where the inner
ring is fixed and the outer free, we suppose that $I_1\rightarrow
\infty$ so that the relative inertia moment is $I=I_2$. Since the
friction torque is $N=\chi \omega$ and the angular velocity
$\omega=v/R$, the friction force is given by
\begin{equation}
F =\frac{N}{R}=\frac{\chi}{R^2} \; v = \frac{\sigma \, A}{\tau} \, v
\label{EQFRICF}
\end{equation}
with the contact area $A=2\pi \ell R$. For nanotubes long enough
$\ell \gg \ell_0$, the relaxation time becomes independent of the
length as shown in Fig. \ref{figLeng} and the shear stress can be
written as $F/A=\sigma v/\tau_{\infty}$. As an example, the shear
stress undergone by a nanotube of radius 10 nm with a rotational
frequency of 1 GHz is approximately 0.047 MPa. This confirms the
expectation that the friction force increases linearly with the
contact area $A$ \cite{Bowden}. Nevertheless, we notice that friction
could be enhanced because of the presence of impurities and
deformations due to the plate fixed to the outer tube in the
experiments of Refs. \cite{Fennimore,Bourlon}.

\begin{figure}[h]
\includegraphics[scale=0.6]{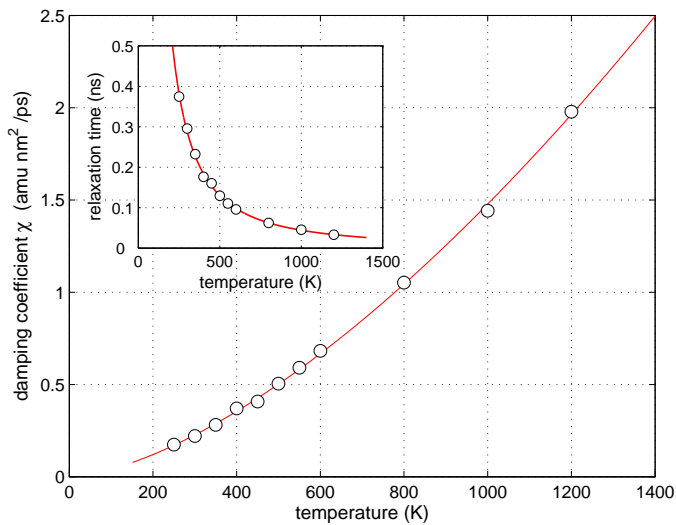}
\caption{(Color online) Dependence of the damping coefficient $\chi$ on temperature.
The circles are the simulation results and the line is a power law
fit of exponent $1.53 \pm 0.04$. The inset is the relaxation time
versus temperature.}
\label{figTemp}
\end{figure}

We have also investigated the dependence of the damping coefficient on the temperature by the method of the autocorrelation function. A
smaller DWNT have been used to gain computation time, namely a (4,4)@(9,9) of length 9.8 \AA. Indeed, the relaxation time is very
large for low temperature. Since the dynamic friction is due to phonon interaction in the DWNT, one expects it to increase with
temperature. On the other hand, at zero temperature, the rotational motion should become ballistic and the diffusion coefficient
$D=k_{\rm B} T /\chi$ should diverge. Indeed, our results confirm these expectations since we find that the damping coefficient depends
on the temperature as the power law $\chi\sim T^{\nu}$ with the exponent $\nu=1.53 \pm 0.04$, as seen in Fig. \ref{figTemp}.
Accordingly, the relaxation time for long enough DWNTs behaves as $\tau_{\infty}=\tau_* (T_*/T)^{\nu}$ where the time $\tau_*$ depends
mainly on the interlayer spacing of the DWNT and $T_*$ is some reference temperature.  We notice that dynamic friction may be non-vanishing at very low temperature where it is dominated by electronic interactions and impurities.

In conclusion, we proposed a stochastic model describing very well the rotational dynamics of DWNTs. We showed that dynamical friction
is proportional to the angular velocity if this later is not too high. Moreover, for long enough nanotubes, the friction force is
proportional to the contact area and can be characterized in terms of a relaxation time $\tau_{\infty}$ which depends on the temperature and the interlayer spacing. For higher velocities, nonMarkovian and nonlinear effects can arise resulting into an increased dissipation.

This research is financially supported by the ''Communaut\'e fran\c
caise de Belgique'' (contract ''Actions de Recherche Concert\'ees''
No. 04/09-312) and the National Fund for Scientific Research
(F.~N.~R.~S. Belgium, contract F. R. F. C. No. 2.4577.04).


\end{document}